\documentclass[12pt]{iopart}

\usepackage[english]{babel}
\usepackage{graphicx}
\usepackage{cite}
\usepackage{color}
\usepackage{bbm}

\begin{document}
	
	\title[]{Simulating para-Fermi oscillators}

	\author{C. Huerta Alderete}
	\address{Instituto Nacional de Astrof\'{\i}sica, \'Optica y Electr\'onica, Calle Luis Enrique Erro No. 1, Sta. Ma. Tonantzintla, Pue. CP 72840, M\'exico}
	
	\author{B. M. Rodr\'iguez-Lara}
	\address{
		Tecnologico de Monterrey, Escuela de Ingenier\'ia y Ciencias, Ave. Eugenio Garza Sada 2501, Monterrey, N.L., M\'exico, 64849.\\ Instituto Nacional de Astrof\'{\i}sica, \'Optica y Electr\'onica, Calle Luis Enrique Erro No. 1, Sta. Ma. Tonantzintla, Pue. CP 72840, M\'exico.}
	\ead{bmlara@itesm.mx}

	\begin{abstract}
		Quantum mechanics allows for a consistent formulation of particles that are neither bosons nor fermions. These para-particles are rather indiscernible in nature. Recently, we showed that strong coupling between a qubit and two field modes is required to simulate even order para-Bose oscillators. Here, we show that finite-dimensional representations of even order para-Fermi oscillators are feasible of quantum simulation under weak coupling. This opens the door to their potential implementation in different contemporaneous quantum electrodynamics platforms. We emphasize the intrinsic value of para-particles for the quantum state engineering of bichromatic field modes. In particular, we demonstrate that binomial two field mode states result from the evolution of para-Fermi vacuum states in the quantum simulation of these oscillators.
	\end{abstract}
	
	
	\maketitle
	
\section{A brief introduction to para-particles}

The harmonic oscillator is an archetype in both classical and quantum mechanics; it can be used to approximate the dynamics of a large number of physical systems and interactions. 
In quantum mechanics, it is straightforward to connect the harmonic oscillator with bosons (fermions) through bilinear commutation (anticommutation) relations\cite{Negele1988p}, 
\begin{eqnarray} 
\left[\hat{b}, \hat{b}^{\dagger} \right]= 1 ~~ \left(\left\{\hat{f}, \hat{f}^{\dagger}\right\} = 1\right),
\end{eqnarray} 
for boson (fermion) annihilation and creation operators,  $\hat{b}$ and $\hat{b}^{\dagger}$ ($\hat{f}$ and $\hat{f}^{\dagger}$).
However, Wigner showed that it is possible to deform these relations leaving the equations of motion unchanged\cite{Wigner1950p711}. A specific deformation was later provided using the reflection operator\cite{Yang1951p788}. In parallel, Green showed that a generalization of the harmonic oscillator yields para-statistics, distributions different from Bose or Fermi statistics\cite{Green1953p270,Greenberg1965pB1155}. In his formulation, the number operator takes a form,
\begin{eqnarray}
\hat{n} = \frac{1}{2} \left\{\hat{b}^{\dagger}, \hat{b} \right\} - \frac{p}{2} ~~ \left(\hat{n} = \frac{1}{2}\left[\hat{f}^{\dagger}, \hat{f}\right] + \frac{p}{2} \right),
\end{eqnarray}
that yield the trilinear commutation relations,
\begin{eqnarray}
\left[ \left\{\hat{b}^{\dagger}, \hat{b} \right\}, \hat{b} \right] =-2 \hat{b} ~~ \left( \left[ \left[ \hat{f}^{\dagger}, \hat{f} \right], \hat{f} \right] =-2 \hat{f} \right), \end{eqnarray} 
of the harmonic oscillator.
This formulation describe the standard Bose and Fermi operators for the statistic order parameter value $p=1$, and so-called ``para-Bose'' (``para-Fermi'') operators for $p > 1$. 
It was later demonstrated that this approach relates to the previous idea of parity deformed oscillators \cite{Yang1951p788,Calogero1969p2191,Vasiliev1991p1115,Plyushchay1997p619} characterized by a deformation parameter equivalent to the statistics order. 
Quantization of these parity deformed oscillators leads to interesting properties \cite{Cusson1969p22,Safonov1991p109,Safonov1994p1195,Wu2002p4506} but their selection rules render their natural occurrence highly unlikely\cite{Hartle1969p2043,Baker2015p929}.
Thus, a method for simulating these para-oscillators is most sought after.

A practical representation of para-particles is found in the parity deformed Heisenberg algebra \cite{Plyushchay1997p619},
\begin{eqnarray}
\left[ \hat{A}, \hat{A}^{\dagger} \right] = 1 + \nu \hat{\Pi}, \quad
\left\{ \hat{\Pi}, \hat{A} \right\} = \left\{ \hat{\Pi}, \hat{A}^{\dagger} \right\}= 0, 
\end{eqnarray}
where the para-particle annihilation (creation) operator is given by $\hat{A}$ ($\hat{A}^{\dagger}$) and the parity operator by $\hat{\Pi}$, such that $\hat{\Pi}^{2} = 1$. 
This algebra characterizes para-Bose (pB) systems of order $p$ when $\nu = p-1$, and para-Fermi (pF) systems of even order $2p$ when $\nu = -(2p+1)$, with $p=1,2,3,\dots$. 	
Standard bosons are recovered when the order is $ p = 1 $, while the lowest order of pF particles recovered is two.
As a consequence, Plyushchay introduced a finite-dimensional deformed $(2p + 1)$-dimensional pF algebra\cite{Plyushchay1997p619},
\begin{eqnarray}
\left[ \hat{I}_{+}, \hat{I}_{-} \right] = 2 \hat{I}_{3} (-1)^{\hat{I}_{3} + p},   \qquad
\left[ \hat{I}_{3}, \hat{I}_{\pm} \right] = \pm \hat{I}_{\pm},
\end{eqnarray}
capable of providing standard fermions, that is the standard representation of $su(2)$, for $p=1$ where $\hat{I}_{3}(-1)^{\hat{I}_{3}+1} = \hat{I}_{3}$.
The latter has a simple relation with the former parity-deformed Heisenberg algebra for $p>1$ because the operators $\left\{\hat{I}_{\pm}, \hat{I}_{3}\right\}$ realize a nonlinear deformation of $su(2)$ involving the parity operator defined as a reflection operator\cite{Plyushchay1997p619}, $\hat{\mathcal{R}} = (-1)^{\hat{I}_{3} + p}$  .

In previous works, we have showed that the two-mode quantum Rabi model (QRM) \cite{Chilingaryan2015p245501,HuertaAlderete2016p414001}, in the homogeneous, strong-coupling limit mimics a collection of even order pB oscillators feasible of quantum simulation in trapped-ions-QED platform \cite{HuertaAlderete2017p013820}. 
Here, we will start from the cross-cavity QRM and show that, in the weak-coupling limit, it might be realized with contemporaneous platforms beyond trapped-ions, for example cavity- and circuit-QED.
Then, we will show the particular partition of its Hilbert space that allows us to describe its dynamics as deformed pF oscillators.  
We will also show that the eigenstates of these deformed pF oscillators are similar to binomial states of the fields via a Schwinger two-boson representation of $SU(2)$.
Finally, we will use this fact to create an educated guess, localized initial-field states, to engineer  two-field mode states through time evolution that produce the collapse and revival of the qubit population inversion without the presence of a coherent initial field state.

\section{The model and its quantum simulation}

\begin{figure}[htbp]
	\centering
	\includegraphics[scale=1]{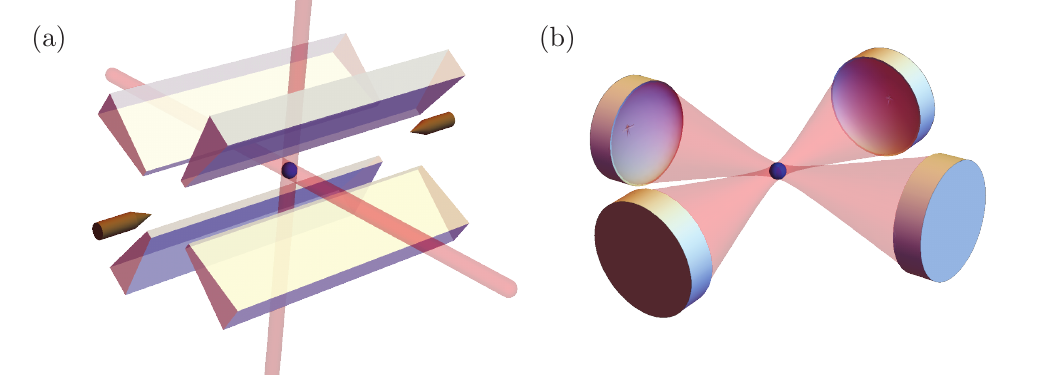}
	\caption{Sketch of the cross-cavity JC model in the  (a) trapped-ion-QED and (b) cavity-QED platforms.}
	\label{fig:Fig1}
\end{figure}

Quantum simulators\cite{Feynman1982p467,Hinds2012p55,Buluta2009p108,Georgescu2014p153} allow us to imitate the dynamics of an \textit{exotic} quantum model in a system that, in principle, is easier to control and measure. 
Within quantum simulation platforms\cite{Buluta2009p108,Wineland1998p147,Blatt2012p277}, trapped ion systems are one of the most important due to the variety of interactions that can be designed\cite{Arrazola2016p30534,Porras2004p207901,Lamata2007p253005,Gerritsma2010p68,Noh2012p033028, Casanova2011p260501,Casanova2012p190502,HuertaAlderete2017p013820}. 
Here, we consider our recent proposal where a trapped ion is driven by two pairs of lasers, each pair orthogonal to the other and tuned to the first side-bands. This system simulates the dynamics of even order pB oscillators under certain model parameters \cite{HuertaAlderete2017p013820}.
This scheme is described by the cross-cavity quantum Rabi model (ccQRM) Hamiltonian  \cite{Chilingaryan2015p245501, HuertaAlderete2016p414001},
\begin{eqnarray}
\hat{H}_{ccQRM} &=& \frac{\omega_{0}}{2} \hat{\sigma}_{3} + \sum_{j}^{2} \omega_{j} \hat{a}^{\dagger}_{j} \hat{a}_{j} + g_{j} \left( \hat{a}^{\dagger}_{j} + \hat{a}_{j}\right) \hat{\sigma}_{j},
\end{eqnarray}
where the two internal levels of an ion interact with two orthogonal vibrational modes with effective coupling strength $ g_{j}$ with $j=1,2$.
The two ion states constitute the effective qubit with transition frequency $\omega_{0}$ and described by Pauli matrices $\hat{\sigma}_j$, with $j=1,2,3$. 
The effective field modes of frecuency $ \omega_{j}$ are described by the creation (annihilation) operators, $ \hat{a}^{\dagger}_{j}$ ( $ \hat{a}_{j}$), such that, $\left[\hat{a}_{j}, \hat{a}^{\dagger}_{k} \right] = \delta_{j,k}$ with $j=1,2$.
When the fields are weakly coupled to the qubit, $g_{j} \ll \omega_{0}$, and near-resonance, $\omega_{j} \sim \omega_{0} $, we can move into a rotating frame defined by the energy of the free system. Then, we can carry out a rotating wave approximation (RWA) to neglect high-frequency terms, and obtain the cross-cavity Jaynes-Cummings (ccJC) model after a  $e^{i \frac{\pi}{2} \hat{a}_{2}^{\dagger} \hat{a}_{2}}$ rotation, 
\begin{equation}\label{eq:ccJC}
\hat{H}_{ccJC} = \delta_1 \hat{a}_{1}^{\dagger} \hat{a}_{1}  + \delta_2 \hat{a}_{2}^{\dagger} \hat{a}_{2} + g_{1}(\hat{a}_{1}^{\dagger}\hat{\sigma}_{-} + \hat{a}_{1}\hat{\sigma}_{+}) +  g_{2}(\hat{a}_{2}^{\dagger}\hat{\sigma}_{-} + \hat{a}_{2}\hat{\sigma}_{+}),
\end{equation}
with detunings $\delta_{j} = \omega_{0} - \omega_{j}$. 
We want to stress that this weak-coupling Hamiltonian can be implemented in our trapped-ion scheme discussed above, sketched in Fig. \ref{fig:Fig1}(a), and in cavity-QED where the qubit is realized by two internal levels of a neutral Rydberg atom coupled to two electromagnetic field modes of orthogonal cavities, Fig. \ref{fig:Fig1}(b). 

\begin{figure}[htbp]
	\centering
	\includegraphics[scale=1]{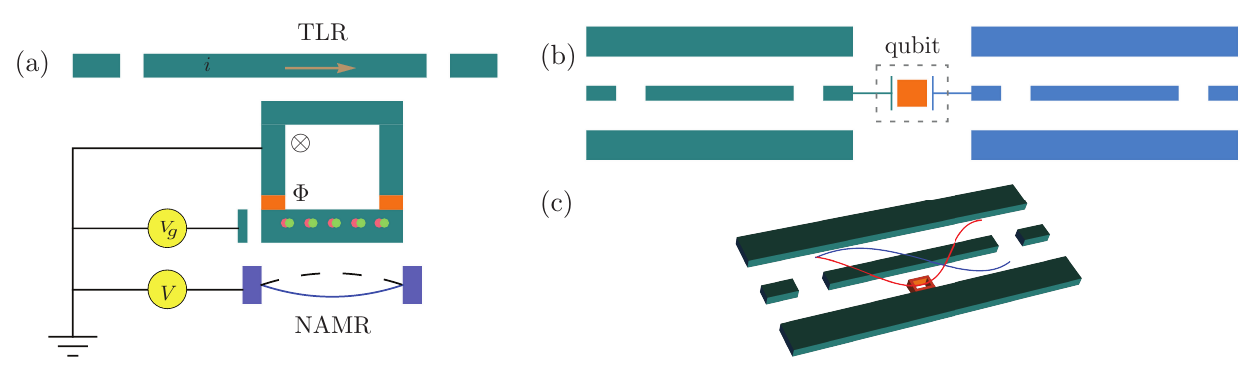}
	\caption{Sketch of ccJC model in circuit QED platform. (a) Mechanical-electrical system controlled by a superconducting qubit, (b) and (c) superconducting transmission lines controlled by a superconductor qubit.}
	\label{fig:Fig2}
\end{figure}

Furthermore, our ccJC Hamiltonian is also feasible in hybrid systems using nanomechanical and transmission line resonators coupled through a quantum node given by a Cooper-pair box or charge qubit, Fig. \ref{fig:Fig2}(a), or two transmission line resonators controlled by a superconducting qubit\cite{Strauch2010p050501,Li2012p014303,Ma2014p062342}, Fig. \ref{fig:Fig2}(b). In addition, an extra rotation to the frame defined by the operator $e^{i \frac{\pi}{2} \hat{a}_{2}^{\dagger} \hat{a}_{2}}$ relates our Hamiltonian to parallel field modes of a coplanar waveguide resonator coupled to an effective superconducting qubit provided by a Cooper-pair box \cite{Moon2005p140504}, charge\cite{Sun2006p022318} or flux qubit \cite{Chen2009p214538}, Fig. \ref{fig:Fig2}(c).

\begin{figure}[htbp]
	\centering
	\includegraphics[scale=1]{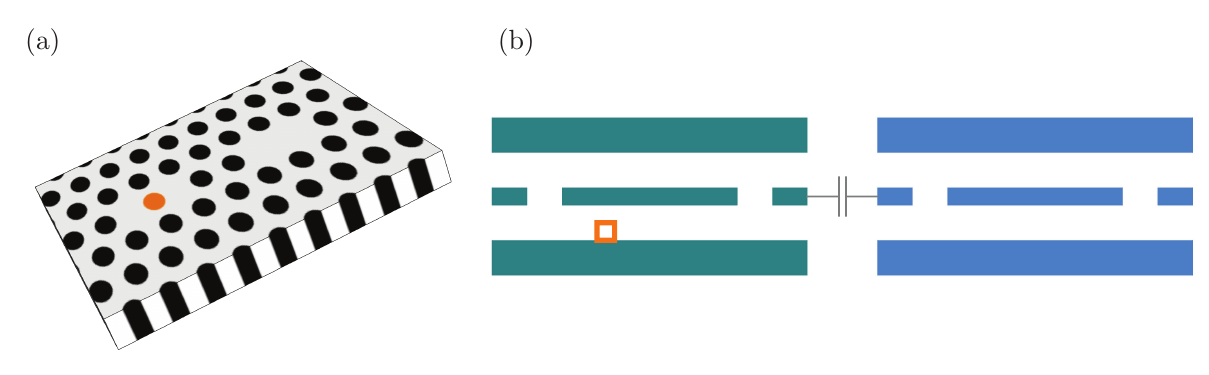}
	\caption{Sketch of $\hat{H}_{D}$ in (a) cavity quantum dot and (b) circuit QED platforms.}
	\label{fig:Fig3}
\end{figure}

We can stop here and notice that a Schwinger two-boson representation of $SU(2)$ might open the door for more  potential experimental realizations. Under an additional rotation,   $e^{i \frac{\pi}{2} \left( \hat{a}_{1}^{\dagger}\hat{a}_{2}-\hat{a}_{1}\hat{a}_{2}^{\dagger}\right)}$, 
the ccJC model can be rewritten in the following form,
\begin{equation}
\hat{H}_{D} = \sum_{j=1}^{2} \Omega_{j} \hat{a}_{j}^{\dagger}\hat{a}_{j} + g(\hat{a}_{1}\hat{\sigma}_{+} + \hat{a}_{1}^{\dagger}\hat{\sigma}_{-}) + \gamma (\hat{a}_{1}^{\dagger}\hat{a}_{2} + \hat{a}_{2}^{\dagger}\hat{a}_{1}),
\end{equation}
where just one boson field is coupled to the qubit under a JC type interaction and the second boson field is coupled to the first one through a beam splitter interaction with modified parameters \cite{HuertaAlderete2016p414001}, $
\Omega_1 = \left( \delta_1 g_{1}^{2} + \delta_2 g_{2}^{2}\right)/{g^{2}}$, $\Omega_2 = \left({\delta_1 g_{2}^{2} + \delta_2 g_{1}^{2}}\right)/{g^{2}}$, $\gamma = (\omega_2 - \omega_1) {g_1 g_2}/{g^{2}}$, and $ g = \sqrt{g_{1}^{2} +g_{2}^{2}}$.
In this frame, our model might be experimentally feasible with coupled photonic-defect resonators including a quantum dot, Fig. \ref{fig:Fig3}(a), or circuit-QED with capacitively-coupled cavities, Fig. \ref{fig:Fig3}(b). In both cases, only one of the cavities is interacting with the effective qubit.
This Hamiltonian, $\hat{H}_{D}$, suggests similar dynamics to that of the single-mode JC model plus a perturbation due to the beam splitter term. 
Considering identical field modes, $\omega_{1}= \omega_{2}$, makes the model solvable.
This simplified version has been widely studied with focus on the description of atomic inversion and generation of two-mode entangled states\cite{Abdalla2002p225,Marchiolli2003p12275,Wildfeuer2003p053801,Larson2006p1867}. Here, we are interested in the general model. 

\section{Partition in orthogonal subspaces}
So far, we have seen that the weak interaction between a two-level system and two boson fields might be realized in several contemporaneous quantum platforms described by QED.
Now, we will show the connection between this model and pF oscillators.
Both our models, $\hat{H}_{ccJC}$ and $\hat{H}_{D}$, conserve the total number of excitations and, therefore, the parity,  $\hat{N}=\hat{a}_{1}^{\dagger} \hat{a}_{1}  + \hat{a}_{2}^{\dagger} \hat{a}_{2} + \frac{1}{2}\left(\hat{\sigma}_{z} + 1\right)$ and $\hat{\Pi}=e^{i \pi \hat{N}}$, respectively, such that $\left[\hat{H}_{x}, \hat{N}\right]= \left[\hat{H}_{x}, \hat{\Pi}\right] = 0$ with $x=ccJC,D$.

The Foulton-Gouterman (FG) approach \cite{Fulton1961p1059,Moroz2016p500004} states that a Hamiltonian of the form $\hat{H} = \hat{A} \otimes \hat{1}_{2} +  \hat{B} \otimes \hat{\sigma}_{x} +  \hat{C} \otimes \hat{\sigma}_{y} +  \hat{D} \otimes \hat{\sigma}_{z}$ can be diagonalized in the qubit basis if there exists an operator $\hat{R}$, such that $\left[\hat{A}, \hat{R}\right] = \left[\hat{B}, \hat{R}\right] = \left\{\hat{C},\hat{R}\right\} = \left\{\hat{D},\hat{R}\right\} = 0$.
The unitary transformation that diagonalizes the Hamiltonian, $\hat{U}_{FG} = \left[\left(\hat{1} + \hat{R} \right) \otimes \left(\hat{\sigma}_{x} + \hat{\sigma}_{z}\right) + \left(\hat{1} - \hat{R} \right) \otimes \left(\hat{1}_{2} - i \hat{\sigma}_{y}\right) \right]/ (2 \sqrt{2})$, usually receives the name of FG transformation.
We can use a $\pi /4$ rotation around $\hat{\sigma}_{y}$ and the FG transformation with the auxiliar operator given by the two-mode parity operator, $\hat{R}=\hat{\Pi}_{12} = e^{i \pi \left( \hat{a}_{1}^{\dagger} \hat{a}_{1}  + \hat{a}_{2}^{\dagger} \hat{a}_{2} \right)}$, to construct a unitary transform, $\hat{U} = \frac{1}{2}\left[(1 - \hat{\Pi}_{12}) \otimes \hat{1} + (1 + \hat{\Pi}_{12}) \otimes \hat{\sigma}_{x}\right]$, that diagonalizes our Hamiltonian in the qubit basis,
\begin{eqnarray}
\hat{H}_{FG} = \hat{U} \hat{H}_{ccJC} \hat{U}^{\dagger} = \hat{H}_{+} \otimes \vert e \rangle\langle e \vert + \hat{H}_{-} \otimes \vert g \rangle\langle g \vert.
\end{eqnarray}
This procedure uncouples the system into two different subspaces, characterized by the two-mode parity-deformed Hamiltonian,
\begin{eqnarray} \label{eq:FG-2JC}
\hat{H}_{\pm} = \sum_{j=1}^{2} \delta_{j} \hat{a}_{j}^{\dagger} \hat{a}_{j} + \frac{g_{j}}{2} \left[\hat{a}_{j}^{\dagger}\left(1 \mp \hat{\Pi}_{12}\right) + \hat{a}_{j}\left(1 \pm \hat{\Pi}_{12}\right)\right].
\end{eqnarray}
In this frame of reference, the total number of excitations in each subspace is also conserved and given by the expression $\hat{N}_{\pm}= \hat{a}_{1}^{\dagger} \hat{a}_{1}  + \hat{a}_{2}^{\dagger} \hat{a}_{2} + \frac{1}{2}\left( 1 \mp \hat{\Pi}_{12}\right)$. 
The conservation of the excitation number allows us to partition the even and odd parity Hilbert subspaces,
\begin{eqnarray}
\mathcal{H}_{+} = \bigoplus^{\infty}_{k=0} \mathcal{H}_{2k}, \quad \mathcal{H}_{-} = \bigoplus^{\infty}_{k=0} \mathcal{H}_{2k+1},
\end{eqnarray}
associated to each one of the two-mode parity-deformed Hamiltonians, $\hat{H}_{\pm}$, into subspaces of dimension $(2 \lambda + 1) $,
\begin{eqnarray}
& \mathcal{H}_{\lambda} = \left\{ \vert \lambda; m \rangle ~ \parallel ~ \vert \lambda; m \rangle\equiv \vert  h(\lambda + m), h( \lambda - m) \rangle \right\},
\end{eqnarray}
span by the vectors $\vert \lambda; m \rangle$ with $ m = -\lambda,-\lambda+1, \dots, 0, \dots, \lambda-1, \lambda$ and the generating function,
\begin{eqnarray}
h(k) = \frac{1}{4} \left( 2k - 1 + e^{i\pi k} \right),
\end{eqnarray}
where the constant mean excitation number in each subspace is given by the parameter $\lambda = 0,1,2,3, \ldots$; even (odd) values of $\lambda$ correspond to subspaces of even (odd) parity $\mathcal{H}_{+}$ ($\mathcal{H}_{-}$).
Henceforth, we will give the name of pF states of even order and dimension $(2\lambda+1)$ to our particular choice of states $\vert \lambda; m \rangle$. 
Before moving forward, we want to show that it is natural to choose this orthogonal basis to partition the Hilbert space of our model.

\begin{figure}[htbp]
	\centering
	\includegraphics[scale=01]{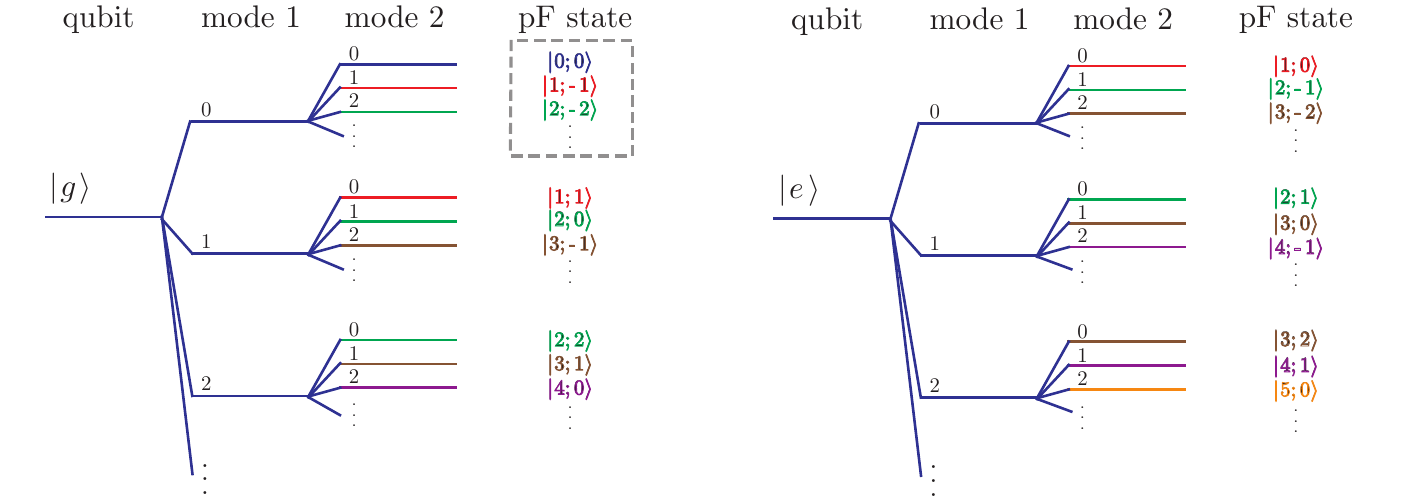}
	\caption{Sketch relating the states of the cross-cavity JC model and the orthonormal pF deformed oscillator basis. The dashed box encloses our choice of lowest energy states, $\vert \lambda; m=-\lambda \rangle$, for each subspace $\mathcal{H}_{\lambda}$.}
	\label{fig:Fig4}
\end{figure}

Our model conserves the total number of excitations and we have used it to label each subspace. For example, the subspace with $\lambda=0$ has dimension one, positive parity, and is spanned by the vector $\vert0; 0 \rangle \equiv \vert g,0,0 \rangle$ equivalent to the qubit being in the ground state and both field modes in the vacuum state, shown in blue in Fig. \ref{fig:Fig4}. The subspace with $\lambda = 1$ has dimension three, negative parity, and the single excitation is either in the qubit or one of the field modes, these states are shown in red in Fig. \ref{fig:Fig4}. The subspace with $\lambda=2$ has dimension five, positive parity, and the vectors spanning it are shown in green in Fig. \ref{fig:Fig4}, and so on.
We chose this representation to have the state with the lowest possible value of the parameter $m$ for a subspace with dimension $\lambda$, that is $m = - \lambda$, given in terms of the ground state of the qubit, the vacuum state of the first mode, and the second mode in a number state with excitation number equal to $\lambda$, as shown by the dashed box in Fig. \ref{fig:Fig4}.

\section{Deformed para-Fermi algebra}
In order to show that our states are pF states, we can project the auxiliary field Hamiltonians, $\hat{H}_{\pm}$, using these bases,
\begin{eqnarray}\label{eq:H2JCeven}
\hat{H}_{\lambda} = \varepsilon_{+} \left\{ \lambda - \frac{1}{2} \left[ 1 - (-1)^{\lambda} \hat{\mathcal{R}} \right]\right\} + \varepsilon_{-} \hat{I}_{3} + \gamma_{+} \left[\hat{I}_{+} + \hat{I}_{-}\right] - \gamma_{-} \left[\hat{I}_{+} - \hat{I}_{-}\right]\hat{\mathcal{R}}, \nonumber \\
\end{eqnarray}
where the effective frequencies are defined as $ \varepsilon_{\pm} = \frac{1}{2}  \left(\delta_{1} \pm \delta_{2}\right) $ and $\gamma_{\pm} = 2^{-3/2} \left(g_{1} \pm g_{2}\right)$. The effective operators,
\begin{eqnarray}
\hat{I}_{3} &=& \hat{a}^{\dagger}_{1} \hat{a}_{1} - \hat{a}^{\dagger}_{2} \hat{a}_{2} , \nonumber \\
\hat{I}_{+} &=& \frac{1}{\sqrt{2}} \left\{ \hat{a}^{\dagger}_{1} \left[1 - (-1)^{\lambda} \hat{\Pi}_{12} \right] + \hat{a}_{2} \left[ 1 +(-1)^{\lambda} \hat{\Pi}_{12} \right] \right\},  \\
\hat{I}_{-} &=& \frac{1}{\sqrt{2}} \left\{ \hat{a}_{1} \left[ 1 + (-1)^{\lambda} \hat{\Pi}_{12} \right] + \hat{a}^{\dagger}_{2} \left[ 1 -(-1)^{\lambda} \hat{\Pi}_{12} \right] \right\}, \nonumber \\
\hat{\mathcal{R}} &=& e^{i \pi (\hat{I}_{3}+ \lambda)}. \nonumber
\end{eqnarray}
realize the deformed pF algebra introduced by Plyushchay\cite{Plyushchay1997p619} in each of the subspaces with constant excitation number.
Furthermore, we can calculate the action of the creation and annihilation operators over the lowest energy state of each subspace,
\begin{eqnarray}
\hat{I}_{-} \hat{I}_{+} \vert \lambda ; -\lambda \rangle = 2\lambda \vert \lambda ; - \lambda  \rangle ,
\end{eqnarray}
and realize that our basis states are pF states\cite{Green1953p270,Greenberg1965pB1155,Plyushchay1997p619} of even order $ p= 2 \lambda$.
The single element $ \vert 0; 0 \rangle $ of the subspace $ \mathcal{H}_{0}$ does not evolve, so the lowest pF order that we can simulate is $p=2$ if we stay inside the subspace $\mathcal{H}_{1}$.
Thus, our model is a quantum simulator of even-order pF oscillators and standard fermions are not covered.

It is worth mentioning that we can give an expression for the population inversion in the laboratory frame, $\hat{\sigma}_{z}$, in terms of the pF frame operators, 
\begin{eqnarray}
\hat{\sigma}_{z} = \left\{ \hat{I}_{+}, \hat{I}_{-}\right\} - (2\lambda + 1) .
\end{eqnarray}
Thereby, it is possible to relate the pF frame evolution to that in the laboratory frame without the need of complicated transformations.
The dynamics of the population inversion can serve as a witness for the dynamics in the pF frame.

\section*{Discussion}

We now turn to the dynamics of our model. 
For the sake of simplicity, we will focus on the evolution of an initial state equal to the pF state $\vert \lambda; -\lambda\rangle$, for identical field modes on resonance with the transition frequency of the qubit,  $\omega_{0}=\omega_{1}=\omega_{2}=\omega$.
This allows us to focus on just the interaction part of our deformed pF oscillators,
\begin{eqnarray}
\hat{H}_{I} = \gamma_{+} \left[\hat{I}_{+} + \hat{I}_{-}\right] - \gamma_{-} \left[\hat{I}_{+} - \hat{I}_{-}\right]\hat{\mathcal{R}},
\end{eqnarray}
and provide a closed form evolution for the lowest energy state in each subspace,
\begin{eqnarray}
\vert \Psi (t) \rangle &=& - i \sum_{k=0}^{\lambda} \sum_{p=0}^{\lambda-k-1} \sum_{q=0}^{k}  \frac{(-1)^{q}}{2^{\lambda}} {\lambda-k-1 \choose p} {k \choose q}  \left[{\lambda \choose k} {\lambda-1 \choose k} {\lambda-1 \choose p+q}^{-1}\right]^{1/2}  \times \nonumber \\
&& \qquad \times   \sqrt{2}\sin \left[ gt \sqrt{2(\lambda-k)} \right] ~~ \vert \lambda, \lambda-1-2(p+q) \rangle \nonumber \\
\nonumber \\
&& + \sum_{k=0}^{\lambda} \sum_{r=0}^{\lambda-k} \sum_{s=0}^{k} \frac{(-1)^{s}}{2^{\lambda}} {\lambda \choose k} {\lambda-k \choose r} {k \choose s} \left[{\lambda \choose r+s}\right]^{-1/2}   \times \nonumber \\
&& \qquad \times \cos \left[ gt \sqrt{2 (\lambda-k) } \right] ~~\vert \lambda, \lambda - 2(r+s) \rangle.
\end{eqnarray}
The evolution of the pF state $\vert \lambda; -\lambda\rangle$ is interesting because it is straightforward to see that this state corresponds to a binomial state with $\eta=1/2$,
\begin{eqnarray}
\vert\Psi(0) \rangle_{D} &=& e^{i \frac{\pi}{2} \left( \hat{a}_{1}^{\dagger}\hat{a}_{2}-\hat{a}_{1}\hat{a}_{2}^{\dagger}\right)} \hat{U}^{\dagger} \vert \lambda, -\lambda \rangle, \nonumber \\
&=& \sum_{k=0}^{\lambda} \left[{\lambda \choose k} (1-\eta)^{\lambda-k} \eta^{k}  \right]^{1/2} \vert \lambda-k,k\rangle \otimes \hat{\sigma}_{x}^{\lambda} \vert e \rangle ,
\end{eqnarray}
in the frame provided by the Hamiltonian in the Schwinger two-boson representation of $SU(2)$.
To the best of our knowledge, this is the first proposal that realizes binomial states since their theoretical introduction \cite{Stoler1985p345}.

\begin{figure}[htbp]
	\centering
	\includegraphics[scale=1]{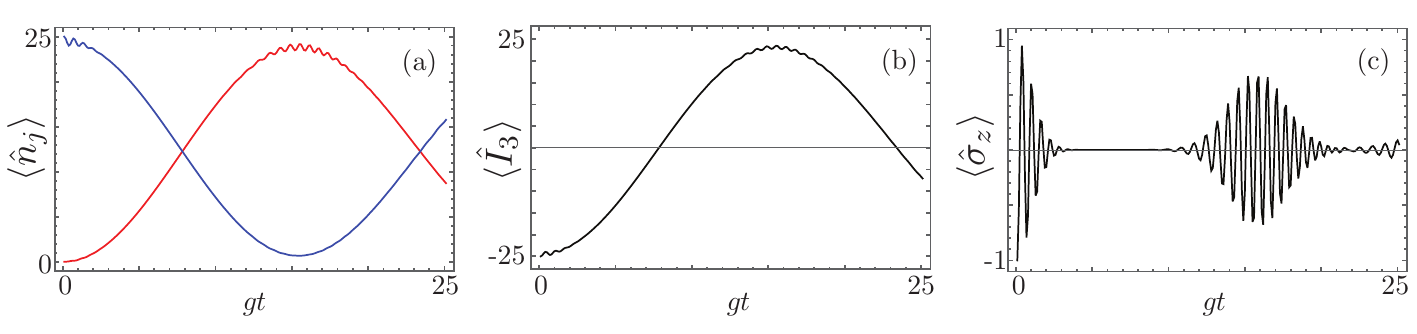}
	\caption{Time evolution for the (a) mean photon number of the first (second) field mode, $\langle \hat{n}_{1(2)} \rangle$, in blue (red), (b) mean deformed pF number, $\langle \hat{I}_{3} \rangle$, and (c) mean population inversion, $\langle \hat{\sigma}_{z}\rangle$, in the laboratory frame for a ccJC model with initial state $\vert g,0, \lambda\rangle $ with parameters $\lambda = 25$, $\omega_{1}=\omega_{2}=\omega_{0}$ and $g_{1}=g_{2}=10^{-3} \omega_{0}$.}
	\label{fig:Fig5}
\end{figure}

The evolution of the pF state $\vert \lambda; -\lambda \rangle$ is equivalent to considering an initial state where the second field mode is in a Fock state with $\lambda $ excitations in it, Fig. \ref{fig:Fig4}, while the first field mode and the qubit are in the vacuum and ground states each. 
In the laboratory frame, the mean photon number evolution of the field modes, under resonant and homogeneous coupling conditions, shows slow excitation exchange with fast perturbation, Fig. \ref{fig:Fig5}(a). 
This behavior stems from the evolution of the mean pF number in the deformed oscillator frame, Fig. \ref{fig:Fig5}(b). 
The two-level system provides the excitation exchange between the field modes.
Thus, its population inversion undergoes Rabi oscillations that collapse and then revive partially,  Fig. \ref{fig:Fig5}(c). 
Here, the lack of a complete revival in the population inversion signals the partial exchange of excitations between the field modes. 
One is reminded of the obvious analogy with the collapse and revival process in the simple Jaynes-Cummings model for an initial coherent state \cite{Narozhny1981p236}.
Furthermore, the revival time for our dynamics has a similar form, $t_{r} = \pi \sqrt{\lambda}/g$, to that found in the standard JC model for initial coherent states \cite{Messina2003p1,Joshi1987p1421}. 
One may wonder about these similitudes. 
Well, the dynamics under these localized initial states allows us to identify the field mode as a type of binomial state. 
It is possible to reduce binomial states to number or coherent states in special limits \cite{Stoler1985p345,VidiellaBarranco1994p5233}.
This can be seen more easily in the Schwinger reference frame, $\hat{H}_{D}$, where the field modes uncouple for resonant frequencies, and we are left with a JC model whose initial field mode state is a binomial state.
In particular, a binomial state with a large mean-excitation number $\lambda$ approximates a coherent state with amplitude $\vert \alpha \vert \approx \sqrt{\lambda}$.
Thus in the Schwinger reference frame, on-resonance and large initial mean-excitation number, we approximate the Jaynes-Cummings model with an initial coherent field that yields the collapse and revival in the dynamics of the population inversion.

\begin{figure}[htbp]
	\centering
	\includegraphics[scale=1]{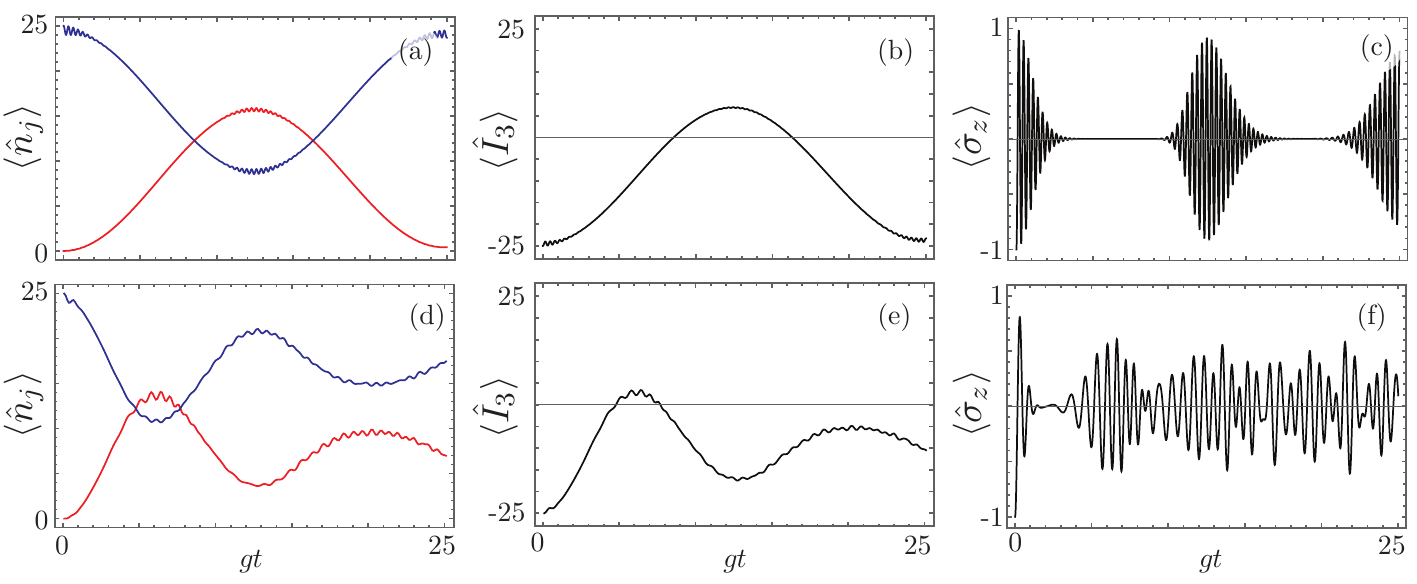}
	\caption{Time evolution for the (a),(d) mean photon number of the first (second) field mode, $\langle \hat{n}_{1(2)} \rangle$, in blue (red), (b),(e) mean deformed pF number, $\langle \hat{I}_{3} \rangle$, and (c),(f) mean population inversion, $\langle \hat{\sigma}_{z}\rangle$, in the laboratory frame for a ccJC model with initial state $\vert g,0, \lambda\rangle $ with parameters $\lambda = 25$, $\omega_{1}=\omega_{2}=\omega_{0}$, (a)-(c) $g_{1} = 2 g_{2} = 10^{-3} \omega_{0}$, and (d)-(f) $2 g_{1} = g_{2} = 10^{-3} \omega_{0}$.}
	\label{fig:Fig6}
\end{figure}

The collapse and revivals in the population inversion are not lost if we break the coupling symmetry, Fig. \ref{fig:Fig6}.
Actually, stronger revivals and extra revival series can be observed for particular coupling ratios, Fig. \ref{fig:Fig6}(c), related to a reduced excitation exchange, Fig. \ref{fig:Fig6}(a), between the field modes when compared to the on-resonance homogeneously coupled case.
This translates into incomplete pF state transfer, Fig. \ref{fig:Fig6}(b).
Furthermore, inhomogeneous couplings can be used to suppress the revival time, Fig. \ref{fig:Fig6}(f), and localize the mean pF number, Fig. \ref{fig:Fig6}(e), which is equivalent to have asymmetric field modes with different mean photon number, Fig. \ref{fig:Fig6}(d), due to the asymmetric coupling between the field modes and the two-level system.

\begin{figure}[htbp]
	\centering
	\includegraphics[scale=1]{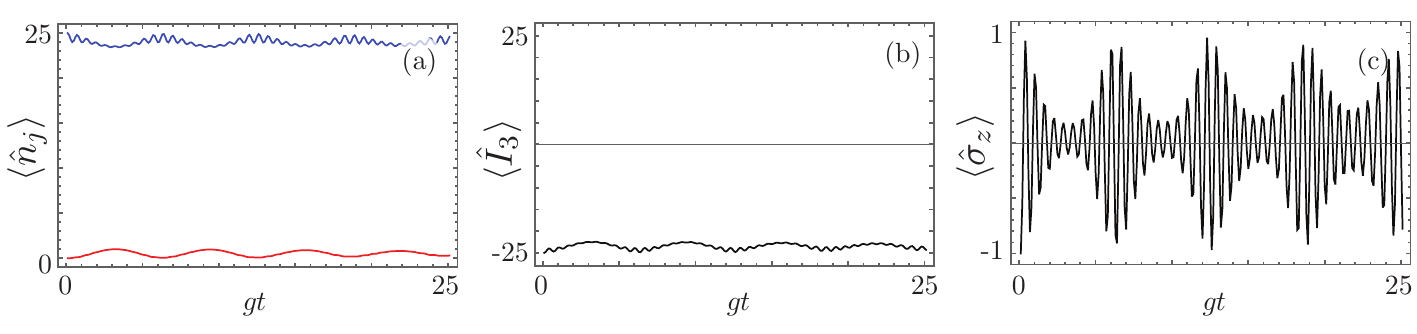}
	\caption{Time evolution for the (a) mean photon number of the first (second) field mode, $\langle \hat{n}_{1(2)} \rangle$, in blue (red), (b) mean deformed pF number, $\langle \hat{I}_{3} \rangle$, and (c) mean population inversion, $\langle \hat{\sigma}_{z}\rangle$, in the laboratory frame for a ccJC model with initial state $\vert g,0, \lambda\rangle $ with parameters $\lambda = 25$, $\omega_{1}=\omega_{0}$ and $\omega_{2}= 1.001\omega_{0}$, (a)-(c) $g_{1} = g_{2} = 10^{-3} \omega_{0}$.}
	\label{fig:Fig7}
\end{figure}

On the other hand, detuning between the two-level system and the field modes can severely impair excitation exchange between the field modes, Fig. \ref{fig:Fig7}(a), leading to highly localized oscillations of the pF state, Fig. \ref{fig:Fig7}(b), accompanied by almost complete revivals of the population inversion, Fig. \ref{fig:Fig7}(c).

\section*{Conclusion}

In summary, we showed that the cross-cavity quantum Rabi model in the weak coupling regime can be described as a collection of isolated parity deformed pF oscillators of even order.
The weak coupling requirement between each field mode and the two-level system opens the door for feasible and highly controllable experimental realizations in trapped-ion-, cavity-, circuit-, and photonic-QED platforms.
Our approach facilitates realizing, for example, the engineering of two-mode binomial states that, to the best of our knowledge, had only been discussed theoretically without relation to an experimental realization.
In addition, the population inversion of the two-level system in the laboratory frame might act as a witness for the two-mode states.
This state engineering of bichromatic field modes is just an example of the uses that might arise from the simulation of para-particles in quantum electrodynamics platforms.

	\section*{Acknowledgments}
	C.H.A. acknowledges financial support from CONACYT doctoral Grant No. 455378 and B.M.R.-L. from CONACYT No. CB-2015-01/255230

	\section*{References}
	

\begin{thebibliography}{45}

\bibitem{Negele1988p}
J.~W.~Negele, and H.~ Orland.
\newblock Quantum many-particle systems.
\newblock (Westview, 1988).

\bibitem{Wigner1950p711}
E.~ P.~ Wigner.
\newblock Do the Equations of Motion Determine the Quantum Mechanical Commutation Relations?
\newblock {\em Phys. Rev.}, {\bf 77}, 711--712 (1950).

\bibitem{Yang1951p788}
L.~ M.~ Yang. 
\newblock A Note on the Quantum Rule of the Harmonic Oscillator.
\newblock {\em Phys. Rev.}, {\bf 84}, 788--790 (1951).

\bibitem{Green1953p270}
H.~S.~Green.
\newblock A generalized method of field quantization.
\newblock {\em Phys. Rev.}, {\bf 90}, 270--273 (1953).

\bibitem{Greenberg1965pB1155}
O.~W.~ Greenberg, and A.~M.~J.~ Messiah.
\newblock Selection rules for parafields and the absence of para particles in nature.
\newblock {\em Phys. Rev.}, {\bf 138}, B1155--B1167 (1965).

\bibitem{Calogero1969p2191}
F.~Calogero.
\newblock Solution of a three-body problem in one dimension.
\newblock {\em J. Math. Phys.}, {\bf 10}, 2191--2196 (1969).

\bibitem{Vasiliev1991p1115}
M.~Vasiliev.
\newblock Higher spin algebras and quantization on the sphere and hyperboloid.
\newblock {\em Int. J. Mod. Phys. A}, {\bf 06}, 1115--1135 (1991).

\bibitem{Plyushchay1997p619}
M.~S.~Plyushchay.
\newblock Deformed Heisenberg algebra with reflection.
\newblock {\em Nucl. Phys. B}, {\bf 491}, 619--634 (1997).

\bibitem{Cusson1969p22}
R.~Cusson.
\newblock Examples of parastatistics.
\newblock {\em Annals of Phys.}, {\bf 55}, 22--40 (1969).

\bibitem{Safonov1991p109}
V.~L.~Safonov.
\newblock On a concept of quasiparticles with parastatistics.
\newblock {\em Phys. Status Solidi (B)}, {\bf 167}, 109--114 (1991).

\bibitem{Safonov1994p1195}
V.~L.~Safonov, and A.~V.~Rozhkov.
\newblock Fr\"{o}hlich's one-dimensional superconductor with para-Fermi statistics.
\newblock {\em Mod. Phys. Lett. B}, {\bf 08}, 1195--1200 (1994).

\bibitem{Wu2002p4506}
L.-A. Wu, and D.~A.~Lidar.
\newblock Qubits as parafermions.
\newblock {\em J. Math. Phys.}, {\bf 43}, 4506--4525 (2002).

\bibitem{Hartle1969p2043}
J.~B.~Hartle, and J.~R.~Taylor.
\newblock Quantum mechanics of paraparticles.
\newblock {\em Phys. Rev.}, {\bf 178}, 2043--2051 (1969).

\bibitem{Baker2015p929}
D.~J.~Baker, H.~Halvorson,  and N.~Swanson.
\newblock The conventionality of parastatistics.
\newblock {\em The Br. J. for Philos. Sci.}, {\bf 66}, 929--976 (2015).

\bibitem{Chilingaryan2015p245501}
S. Chilingaryan, and B.~M.~Rodr{\'\i}guez-Lara.
\newblock Exceptional solutions in two-mode quantum Rabi models.
\newblock {\em J. Phys. B: At. Mol. Opt. Phys.}, {\bf 48}, 245501 (2015).

\bibitem{HuertaAlderete2016p414001}
C. Huerta~Alderete, and B.~M.~Rodr{\'\i}guez-Lara.
\newblock Cross-cavity quantum Rabi model.
\newblock {\em J. Phys. A: Math. Theor.}, {\bf 49}, 414001 (2016).

\bibitem{HuertaAlderete2017p013820}
C. Huerta~Alderete, and B.~M.~Rodr\'{\i}guez-Lara.
\newblock Quantum simulation of driven para-Bose oscillators.
\newblock {\em Phys. Rev. A} {\bf 95}, 013820 (2017).

\bibitem{Feynman1982p467}
R.~P.~Feynman.
\newblock Simulating physics with computers.
\newblock {\em Int. J. Theor. Phys.}, {\bf 21}, 467--488 (1982).

\bibitem{Hinds2012p55}
E.~Hinds, and R.~Blatt.
\newblock Manipulating individual quantum systems.
\newblock {\em Nat.}, {\bf 492}, 55 (2012).

\bibitem{Buluta2009p108}
I.~Buluta, and F.~Nori.
\newblock  Quantum simulators.
\newblock {\em Sci}, {\bf 326}, 108--111 (2009).

\bibitem{Georgescu2014p153}
I.~M.~Georgescu,  S.~Ashhab, and F.~Nori.
\newblock Quantum simulation.
\newblock {\em Rev. Mod. Phys.}, {\bf 86}, 153--185 (2014).

\bibitem{Wineland1998p147}
D.~J.~Wineland, \emph{et~al.}
\newblock Trapped-ion quantum simulator.
\newblock {\em Phys. Scripta}, {\bf 1998}, 147 (1998).

\bibitem{Blatt2012p277}
R.~Blatt, and C.~F.~Roos.
\newblock Quantum simulations with trapped ions.
\newblock {\em Nat. Phys.}. {\bf 8}, 277--284 (2012).

\bibitem{Arrazola2016p30534}
I.~Arrazola,and J.~S.~Pedernales, and L.~Lamata, and E.~Solano.
\newblock Digital-analog quantum simulation of spin models in trapped ions.
\newblock {\em Sci. Rep.}, {\bf 6}, 30534 (2016).

\bibitem{Porras2004p207901}
D.~Porras, J.~I.~Cirac.
\newblock Effective quantum spin systems with trapped ions.
\newblock {\em Phys. Rev. Lett.}, {\bf 92}, 207901 (2004).

\bibitem{Lamata2007p253005}
L.~Lamata, J. Le\'on, T. Sch\"atz, and E. Solano.
\newblock Dirac equation and quantum relativistic effects in a single trapped ion.
\newblock {\em Phys. Rev. Lett.}, {\bf 98}, 253005 (2007).

\bibitem{Gerritsma2010p68}
R.~Gerritsma, \emph{et~al.}
\newblock Quantum simulation of the Dirac equation.
\newblock {\em Nat.}, {\bf 463}, 68--71 (2010).

\bibitem{Noh2012p033028}
C.~Noh, B.~M.~Rodríguez-Lara, and D.~G.~Angelakis.
\newblock Quantum simulation of neutrino oscillations with trapped ions.
\newblock {\em New J. Phys.}, {\bf 14}, 033028 (2012).

\bibitem{Casanova2011p260501}
J.~Casanova, \emph{et~al.}
\newblock Quantum simulation of quantum field theories in trapped ions.
\newblock {\em Phys. Rev. Lett.}, {\bf 107}, 260501 (2011).

\bibitem{Casanova2012p190502}
J.~Casanova, A.~Mezzacapo, L.~Lamata, and E.~Solano.
\newblock Quantum simulation of interacting fermion lattice models in trapped ions.
\newblock {\em Phys. Rev. Lett.}, {\bf 108}, 190502 (2012).

\bibitem{Strauch2010p050501}
F.~W.~Strauch, K.~Jacobs, and R.~W.~Simmonds.
\newblock Arbitrary control of entanglement between two superconducting resonators.
\newblock {\em Phys. Rev. Lett.}, {\bf 105}, 050501 (2010).

\bibitem{Li2012p014303}
P.-B.~Li, S.-Y.~Gao, and F.-L.~Li.
\newblock Engineering two-mode continuous-variable entangled states of distant atomic spin ensembles with superconducting quantum circuits.
\newblock {\em Phys. Rev. A}, {\bf 85}, 014303 (2012).

\bibitem{Ma2014p062342}
S.-L.~Ma, \emph{et~al.}
\newblock Controllable generation of two-mode-entangled states in two-resonator circuit QED with a single gap-tunable superconducting qubit.
\newblock {\em Phys. Rev. A}, {\bf 90}, 062342 (2014).


\bibitem{Moon2005p140504}
K.~Moon, and S.~M.~Girvin.
\newblock Theory of microwave parametric down-conversion and squeezing using circuit QED.
\newblock {\em Phys. Rev. Lett.}, {\bf 95}, 140504 (2005).

\bibitem{Sun2006p022318}
C.~P.~Sun, L.~F.~Wei, Y.-X.~Liu, and F.~Nori.
\newblock Quantum transducers: Integrating transmission lines and nanomechanical resonators via charge qubits.
\newblock {\em Phys. Rev. A}, {\bf 73}, 022318 (2006).

\bibitem{Chen2009p214538}
M.-Y.~Chen, M. W.-Y. Tu,  and W.-M. Zhang.
\newblock Entangling two superconducting $LC$ coherent modes via a superconducting flux qubit.
\newblock {\em Phys. Rev. B}, {\bf 80}, 214538 (2009).

\bibitem{Abdalla2002p225}
M.~S.~Abdalla, M. Abdel-Aty, and A.~S.~Obada.
\newblock Quantum entropy of isotropic coupled oscillators interacting with a single atom.
\newblock {\em Opt. communications}, {\bf 211}, 225--234 (2002).

\bibitem{Marchiolli2003p12275}
M.~A.~Marchiolli, R.~J.~Missori,  and J.~A.~Roversi.
\newblock Qualitative aspects of entanglement in the Jaynes--Cummings model with an external quantum field.
\newblock {\em J. Phys. A: Math. Gen.}, {\bf 36}, 12275 (2003).

\bibitem{Wildfeuer2003p053801}
C.~Wildfeuer, and D.~Schiller.
\newblock Generation of entangled N-photon states in a two-mode Jaynes-Cummings model.
\newblock {\em Phys. Rev. A}, {\bf 67}, 053801 (2003).

\bibitem{Larson2006p1867}
J.~Larson.
\newblock Scheme for generating entangled states of two field modes in a cavity.
\newblock {\em J. Mod. Opt.}, {\bf 53}, 1867--1877 (2006).

\bibitem{Fulton1961p1059}
R.~ L. Fulton,  and M.~ Gouterman.
\newblock Vibronic coupling. I. Mathematical treatment for two electronic states.
\newblock {\em The J. Chem. Phys.}, {\bf 35}, 1059--1071 (1961).

\bibitem{Moroz2016p500004}
A.~Moroz.
\newblock Generalized Rabi models: Diagonalization in the spin subspace and differential operators of Dunkl type.
\newblock {\em EPL (Europhysics Letters)}, {\bf 113}, 50004 (2016).

\bibitem{Stoler1985p345}
D.~Stoler, B.~ Saleh, and M.~ Teich.
\newblock Binomial states of the quantized radiation field.
\newblock {\em Opt. Acta: Int. J. Opt.}, {\bf 32}, 345--355 (1985).

\bibitem{Narozhny1981p236}
N.~B.~ Narozhny,J.~J.~Sanchez-Mondragon, and J.~H.~Eberly.
\newblock Coherence versus incoherence: Collapse and revival in a simple quantum model.
\newblock {\em Phys. Rev. A}, {\bf 23}, 236--247 (1981).

\bibitem{Messina2003p1}
A.~Messina, S.~Maniscalco, and A.~Napoli.
\newblock Interaction of bimodal fields with few-level atoms in cavities and traps.
\newblock {\em J. Mod. Opt.}, {\bf 50}, 1--49 (2003).

\bibitem{Joshi1987p1421}
A.~Joshi, and R.~Puri.
\newblock Effects of the binomial field distribution on collapse and revival phenomena in the Jaynes--Cummings model.
\newblock {\em J. Mod. Opt.}, {\bf 34}, 1421--1431 (1987).

\bibitem{VidiellaBarranco1994p5233}
A.~Vidiella-Barranco, and J.~A.~Roversi.
\newblock Statistical and phase properties of the binomial states of the electromagnetic field.
\newblock {\em Phys. Rev. A}, {\bf 50}, 5233--5241 (1994).

\end{thebibliography}

\end{document}